\pdfoutput=1
\documentclass{aa}  
\usepackage{graphicx}
\usepackage{txfonts}
\usepackage{natbib}
\usepackage{dcolumn}
\usepackage{hyperref}

\newcolumntype{.}{D{.}{.}{-1}}
\newcolumntype{;}{D{;}{.}{7}}
\bibpunct{(}{)}{;}{a}{}{,} % to follow the A&A style

\newcommand{\solm}{M$_{\odot}$\ }

\begin{document}

\authorrunning{Kunneriath, Eckart et al.}
\titlerunning{The Galactic centre mini-spiral in the mm-regime }
\title{The Galactic centre mini-spiral in the mm-regime}
\subtitle{}

\author{D. Kunneriath$^{1,2,3}$,
       A. Eckart$^{1,2}$,
       S. N. Vogel$^{4}$,
       P. Teuben$^{4}$,
       K. Mu\v{z}i\'c$^{5}$,
       R. Sch\"odel$^{6}$,
       M. Garc\'{\i}a-Mar\'{\i}n$^1$,
       J. Moultaka$^{7}$,
       J. Staguhn$^{4,8,9}$,
       C. Straubmeier$^1$,
       J. A. Zensus$^{2,1}$,
       M. Valencia-S.$^{1,2}$,
       V. Karas$^{3}$
}

\offprints{D. Kunneriath (devaky@astro.cas.cz)}

   \institute{ I.Physikalisches Institut, Universit\"at zu K\"oln,
              Z\"ulpicher Str. 77, 50937 K\"oln, Germany
	 \and
             Max-Planck-Institut f\"ur Radioastronomie, 
             Auf dem H\"ugel 69, 53121 Bonn, Germany
         \and
		Astronomical Institute, Academy of Sciences, Bo\v{c}n\'{i}í II 1401, CZ-14100 Prague, Czech Republic
	\and
	 Department of Astronomy, University of Maryland, College Park, MD 20742-2421, USA
	\and
	Department of Astronomy and Astrophysics, University of Toronto, 50 St. George Str., Toronto ON M5S 3H4, Canada
	 \and
  Instituto de Astrof\'isica de Andaluc\'{i}a - CSIC, Glorieta de la Astronom\'{i}a S/N, 18008, Spain
         \and
	  LATT, Universit\'e de Toulouse, CNRS, 14, Avenue Edouard Belin, 31400 Toulouse, France
	\and
	  NASA Goddard Space Flight Ctr., Greenbelt, MD 20771, USA
	\and
	Department of Physics and Astronomy, Johns Hopkins University, Baltimore, MD 21218, USA
              }

\date{Received  / Accepted }

\abstract{ The mini-spiral is a feature of the interstellar medium in the central $\sim$2\,pc of the Galactic center. 
It is composed of several streamers of dust and ionised and atomic gas with temperatures between a few 100\,K to 10$^4$\,K. 
There is evidence that these streamers are related to the so-called circumnuclear disk of molecular gas and are
 ionized by photons from massive, hot stars in the central parsec.
}{                                                                              
We attempt to constrain the emission mechanisms and physical properties of the ionized gas and dust of the mini-spiral region with the 
help of our multiwavelength data sets. 
}{Our observations were carried out at 1.3\,mm and 3\,mm with the mm interferometric array CARMA in 
California in March and April 2009, with the MIR instrument VISIR at ESO's VLT in June 2006, and the NIR Br$\gamma$ with 
VLT NACO in August 2009. 
}{We present high resolution maps of the mini-spiral, and obtain a spectral index of 0.5$\pm$0.25 for Sgr~A*, indicating an 
inverted synchrotron spectrum. We find electron densities within the range 0.8--1.5$\times$10$^{4}$\,cm$^{-3}$ for the mini-spiral 
from the radio continuum maps, along with a dust mass contribution of $\sim$0.25\,\solm from the MIR dust continuum, 
and extinctions ranging from 1.8--3 at 2.16\,$\mu$m in the Br$\gamma$ line. }
{We observe a mixture of negative and positive spectral indices in our 1.3\,mm and 3\,mm observations of the extended emission of 
the mini-spiral, which we interpret as evidence that there are a range of contributions to the thermal free-free emission by the ionized gas emission 
and by dust at 1.3\,mm. }

\keywords{black hole physics, infrared: general, accretion, accretion disks, Galaxy: center, nucleus,
Black Holes: individual: SgrA*}

%\titlerunning{VLT/mm flare from Sgr~A*}
%   \authorrunning{Eckart, Sch\"odel, Garc\'{\i}a-Mar\'{\i}n, Witzel, Weiss  et al.}  
   \maketitle

\section{Introduction}
The radio emission from the inner few parsecs of the Galactic centre (GC) region is dominated by the interstellar 
medium of the  circumnuclear disk (CND) and the mini-spiral, the nuclear stellar 
cluster, and the supermassive black hole ($\sim$4$\times$10$^{6} M_{\sun}$) associated with the radio and infra-red source Sgr A* 
\citep[and references therein]{eckart1996,schoedel2002,ghez2009,genzel2010}. 
The CND is assumed to be an association of clouds/filaments of dense (10$^{4}$-10$^{7}$ cm$^{-3}$) and warm (several 100 K) 
molecular gas ($\sim$10$^5$ M$_{\sun}$ of dust and gas, \citep{christopher2005}), which orbits the nucleus
in a circular rotation pattern, with a sharp inner edge at 1.5 pc that extends no 
further than 7 pc from the centre \citep{guesten1987,christopher2005,montero2009}. The region inside this cavity contains atomic 
and ionized gas. 

Several streamers of infalling gas and dust from the inner edge of the CND form the mini-spiral, 
which consists of four main components: the northern arm, the western arc, the eastern arm and the bar \citep[Labelled in fig. \ref{regular3mm},][]{ekers1983,lo1983,zhao2009}. 
These streams have been modelled as a bundle of three elliptical Keplerian orbits, with some significant deviations 
\citep{zhao2009,zhao2010}, the northern and eastern arms being on highly elliptical orbits, while the western arc is nearly circular. 
The deviations from Keplerian motion have been assumed to be caused by stellar winds. 
The central region is also bright in the near-infrared (NIR) and mid-infrared (MIR) regimes, where the stellar population dominates 
the NIR and warm dust emission accounting for MIR emission at 10\,$\mu$m \citep{becklin1975,becklin1978,viehmann2006}. Analyses of VLA observations 
at 5 GHz \citep{brown1981} and Ne[II] line observations \citep{serabyn1985,serabyn1991} have 
also found peaks of emission coinciding with these 10\,$\mu$m thermal peaks. 

\begin{table*}[!Htbp]
\centering
{\begin{small}
\begin{tabular}{cccll}
\hline
Telescope & Instrument/Array & $\lambda$ & UT and JD & UT and JD \\
 & & & Start Time & Stop Time \\
\hline
\\
CARMA	       & C array   & 3.0~mm	 & 2009 17 May 07:21:23.5 & 17 May 12:24:27.5\\
               &           &             & JD 2454968.806522      & JD  2454969.016985\\
CARMA	       & D array   & 1.3~mm	 & 2009 28 Mar 11:20:03.5 & 28 Mar 14:39:18.5\\
               &           &             & JD 2454918.972263      & JD 2454919.110631 \\
CARMA          & C array   & 1.3~mm      & 2009 19 Apr 10:07:21.5 & 19 Apr 13:13:13.5  \\
               &           &             & JD 2454940.921777      & JD 2454941.050851 \\
VLT            & NACO      & 2.16~$\mu$m & 2009 05 Aug 22:59:14.59& 05 Aug 23:30:44.34\\
               &           &             & JD 2455049.457808      & JD 2455049.479680\\
VLT            & VISIR     & 8.6~$\mu$m  & 2006 05 Jun 04:53:43.77& 05 Jun 10:25:07.49\\
               &           &             & JD   2453891.703979    & JD 2453891.934114\\

\hline
\end{tabular}
\end{small}}
\caption{Log of the mm, NIR, and MIR observations.}
\label{log1}
\end{table*}
The mini-spiral is a region with a complex temperature and density structure with 
temperatures ranging from a few hundred K in the dust component to 10$^4$\,K in the plasma component \citep{moultaka2005}. 
Dust emission modelling of the central parsec region in the infrared (4.8-20.0\,$\mu$m) 
reveals temperatures of a few hundreds K with IRS 7 (1300\,K) and IRS 3 (600\,K) being the 
two hottest sources \citep{gezari1996mirmodel,moultaka2004,moultaka2005}. The source 
Sgr A* is very bright at radio/mm wavelengths at a typical resolution of a few arcseconds. 
Interferometric observations at these wavelengths can separate 
the flux density contributions of Sgr A* and the mini-spiral region from the surrounding 
CND. \cite{zhao2009,zhao2010} presented 1\,mm maps of $\sim$0.15$''$ resolution of the mini-spiral region, 
while \cite{mauerhan2005} presented maps of the highest resolution yet achieved at 3\,mm of $\sim$4$''$ resolution.

The spectral index behaviour of the mini-spiral at these wavelengths has never been studied. 
We therefore embark on such an investigation in the present paper. Our work \citep[the preliminary results of 
which were published in][]{kunneriath2011} complements the discussion of \cite{garcia-marin2011}, 
who presented sub-mm emission maps of the inner 37$\times$34 pc$^{2}$ of the 
GC region, including the CND and surrounding molecular clouds. These authors derived temperature maps of this 
region, which revealed dust temperatures between 10\,K and 20\,K in the CND and molecular clouds. Spectral 
index maps of the same region indicate that the spectral indices range from -0.8 to 1.0 in these low temperature 
regions, which has been interpreted as evidence of a combination of dust, synchrotron, and free-free emission. 
In the regions immediately surrounding Sgr A*, they derived spectral indices of -0.6$<$$\alpha$$<$0.0, which was 
explained as a combination of 70-90\% synchrotron and 10-30\% dust and free-free emission. However, at sub-mm wavelengths, 
the emission of Sgr A* dominates the central few arcseconds, making an analysis of the emission mechanisms of the mini-spiral 
difficult at these wavelengths.

In the radio regime, \cite{ekers1975} published a spectral index map of the Galactic centre region 
between 6\,cm and 20\,cm wavelengths, and obtained a non-thermal spectral index of $-$0.3 for the diffuse emission from Sgr A 
west, and a thermal spectrum for the Sgr A west spiral. However, at 20 cm, the spiral features are not clearly resolved, and 
the emission could be optically thick at $\lambda\ge$6\,cm \citep{mezger1989}.

We present ($\sim$2--4$''$) high spatial resolution maps from observations of the Galactic centre region  at 1.3\, and 3\,mm 
performed in 2009 using 
CARMA, yielding a spectral index map obtained at millimetre frequencies, a comparison to the MIR dust continuum 
and NIR Br$\gamma$ emission in 
the same region, and an analysis of the emission processes in the mini-spiral region. Section 2 is dedicated to a description of 
the observations and data reduction methods, while Sects. 3 and 4 provide the results and summary. Physical properties of the dust 
and gas in selected regions are calculated and tabulated in Table \ref{fluxtable}.   
\begin{figure}[!Htp]
%\begin{minipage}[t]{.4\textwidth}
 \begin{center}
\includegraphics[scale=0.4,angle=-90]{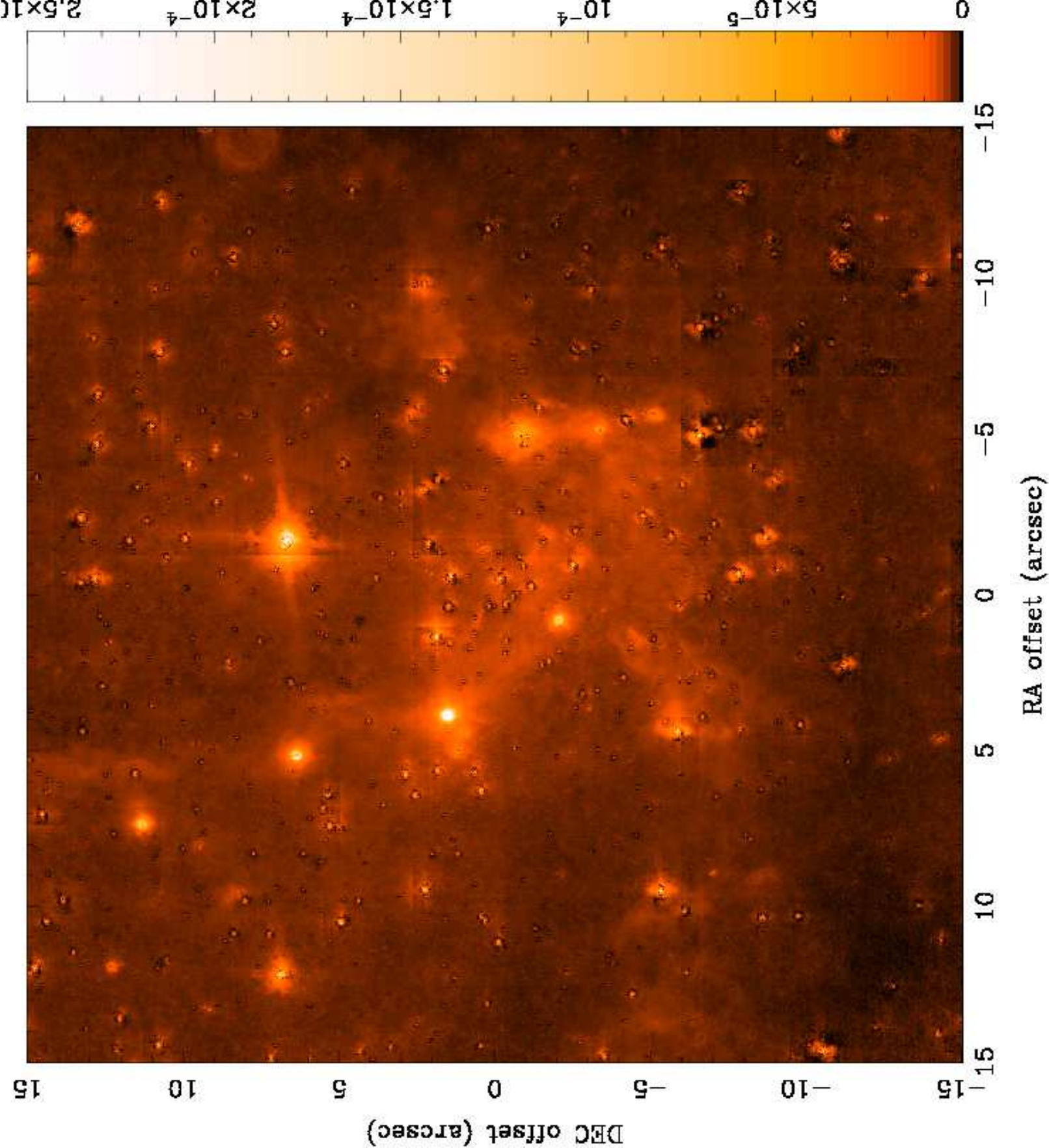}
\caption{\small Br$\gamma$ image of the central 30$''$ mini-spiral region around Sgr~A*. The colour bar indicates flux density units of Jy/pixel.}
\label{fig:brgamma}  
 \end{center}
%\end{minipage}  
\end{figure}

\section{Observations and data reduction} 
\subsection{Millimetre observations}
%\subsubsection{3 mm data}
The Combined Array for Research in mm-wave Astronomy (CARMA)\footnote{Support for CARMA construction was derived from the states of 
California, Illinois, and Maryland, the Gordon and Betty Moore Foundation, the Kenneth T. and 
Eileen L. Norris Foundation, the Associates of the California Institute of Technology, 
and the National Science Foundation.  Ongoing CARMA development and operations are 
supported by the National Science Foundation under a cooperative agreement, and by the 
CARMA partner universities.} is a millimetre array consisting of 
15 telescopes, including six 10.4\,m telescopes with a half-power beamwidth (HPBW) of 69$''$ at 100\,GHz and 30$''$ at 230\,GHz, 
and nine 6.1\,m telescopes 
with HPBW of 115$''$ at 100\,GHz and 50$''$ at 230\,GHz. Five array configurations are possible, ranging from A (0.25--2\,km) 
to E (8--66\,m), and reaching an angular resolution of up to 0.15$''$ at 230\,GHz in the A-array.

The observations at 3\,mm (100\,GHz) were made in May 2009 with CARMA in the C configuration 
with intermittent observations of 3C279 for bandpass calibration and 1733-130 for phase and amplitude calibration. 
The only flagging done was to remove shadowing effects. 
At 1.3\,mm (230\,GHz), observations were made in March and April 2009 in the C and D configuration, 
using MWC349 for bandpass calibration and 1733-130 for phase and amplitude calibration. Owing to anomalous sytem temperature values, 
the complete antenna 7 dataset was flagged and antenna 14 was 
flagged from 11:50.9 to 12:10.0 and 10:00.0 to 11:10.0 UT. The uv-coverages of the C array at 3\,mm and the D array at 1.3\,mm are 
shown in the Appendix. 

Table \ref{log1} gives the details of observations with start and stop times. All observations were centred 
at $\alpha$(J2000.0)=17:45:40.04 and $\delta$(J2000.0)=-29:00:28.09. The data 
reduction was performed with {\it Miriad}, an interferometric data reduction package \citep{miriad1995}. The calibrated 
visibilities were inverted using the {\it mosaic} option and a CLEAN algorithm was applied to produce 
the final radio maps. 
\begin{figure}[!Htbp]
\centering
\includegraphics[scale=0.55,angle=0]{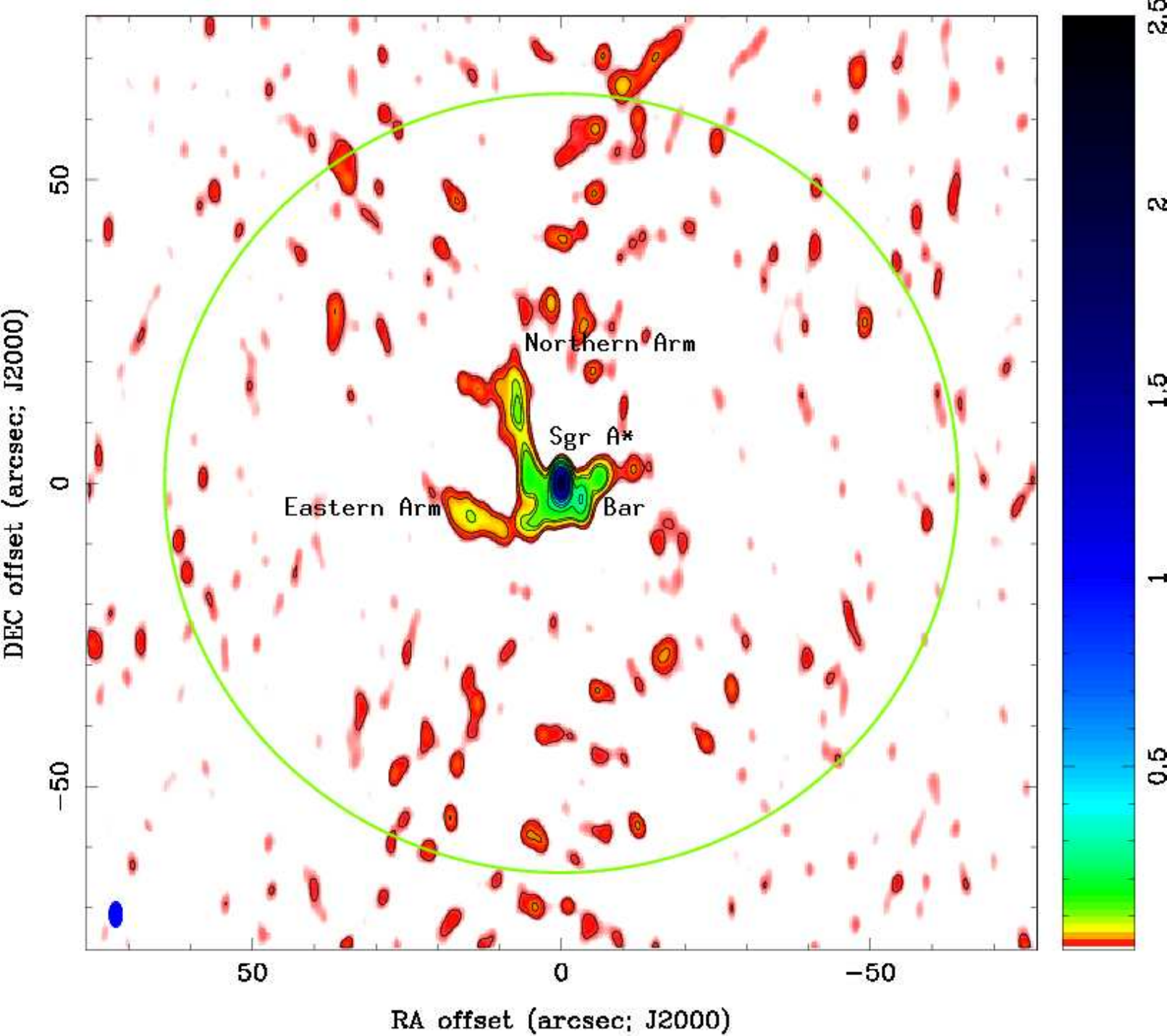}
\caption{\small Three-mm map of the mini-spiral region at resolution 2.68$''$ by 1.71$''$ (P.A=-20.7$^{\circ}$). The green circle marks the HPBW. Contour levels 
are 0.02, 0.03, 0.04, 0.05, 0.1, 0.6, 0.9, 1.2, 1.5, and 2.5 Jy/beam.}
\label{regular3mm} 
\centering
\includegraphics[scale=0.37,angle=0]{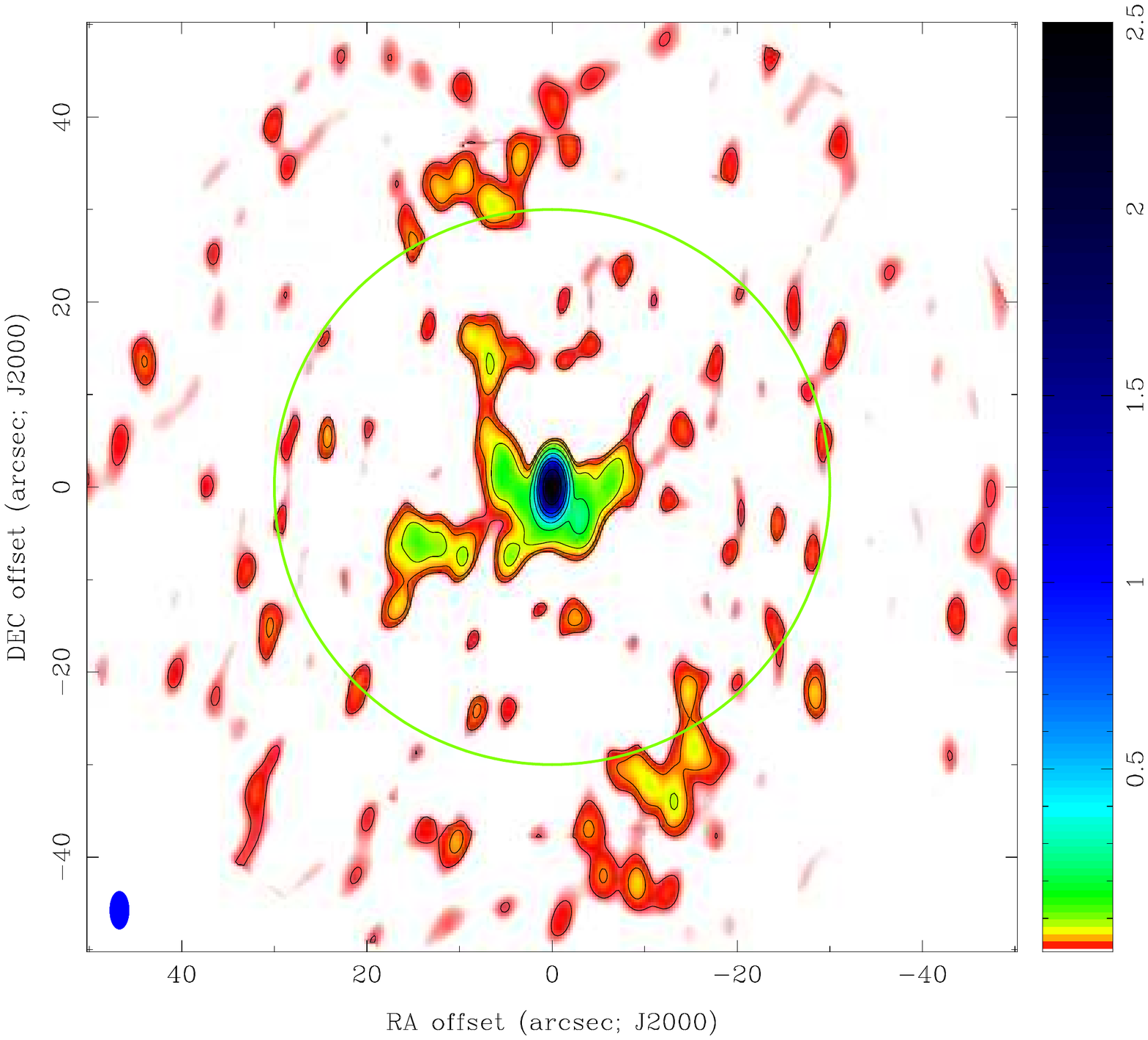}
\caption{\small A 1.3\,mm map of the mini-spiral region at resolution 4.42$''$ by 1.23$''$ (P.A=1.0$^{\circ}$). The green circle marks the HPBW. Contour levels 
are 0.02, 0.03, 0.04, 0.05, 0.1, 0.6, 0.9, 1.2, 1.5, and 2.5 Jy/beam.}
\label{regular1mm}    
\end{figure}

\begin{figure*}[!ht]
\centering
\includegraphics[scale=0.5,angle=0]{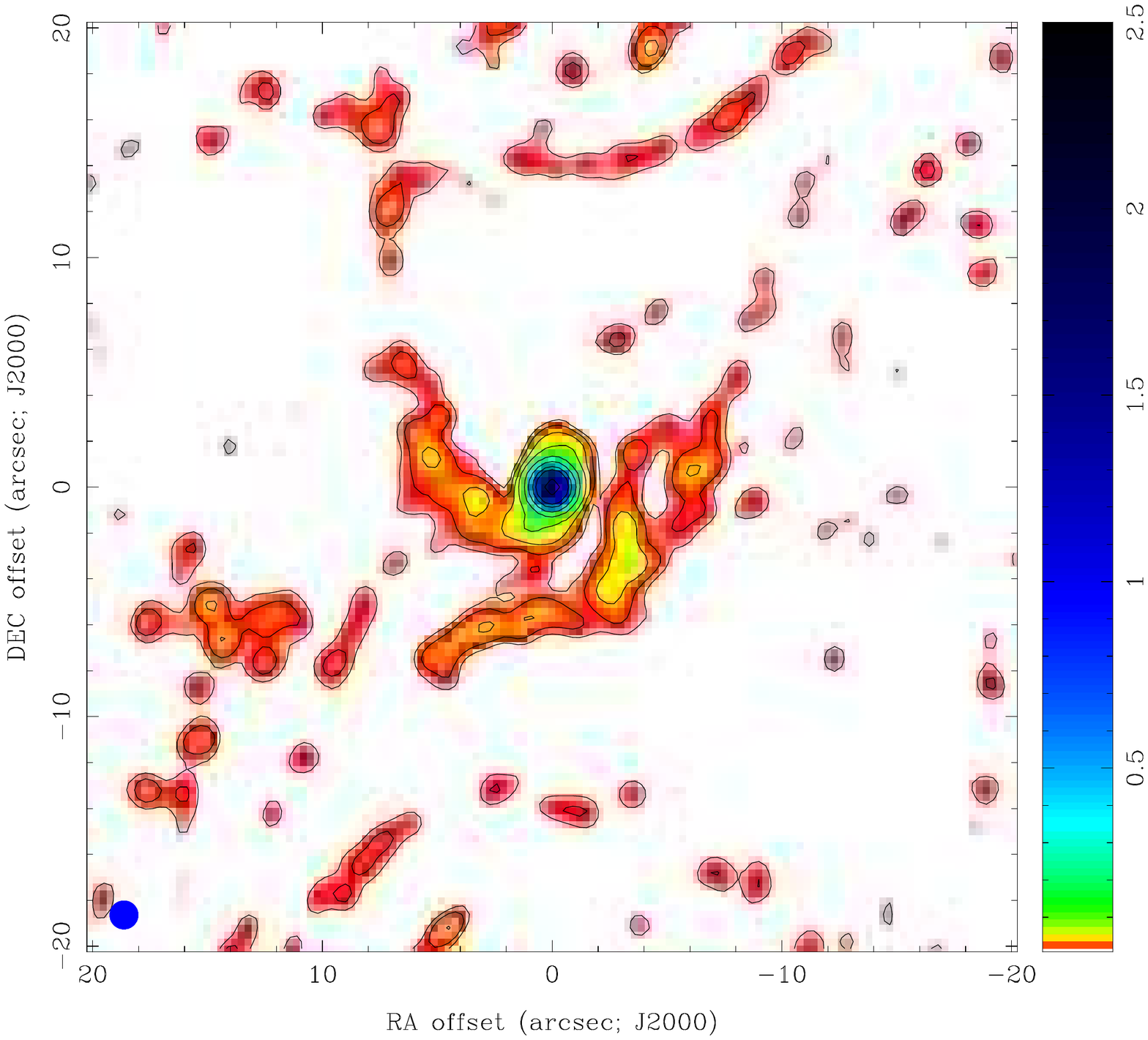}
\caption{\small The mini-spiral region image obtained from 1.3\,mm CD array configuration with a synthesised circular beam size 
of 1.2$''$. Contour levels are 0.015, 0.02, 0.03, 0.04, 0.05, 0.1, 0.6, 0.9, 1.2, 1.5, and 2.5 Jy/beam.}
\label{fig:highres}  
\includegraphics[scale=0.5,angle=0]{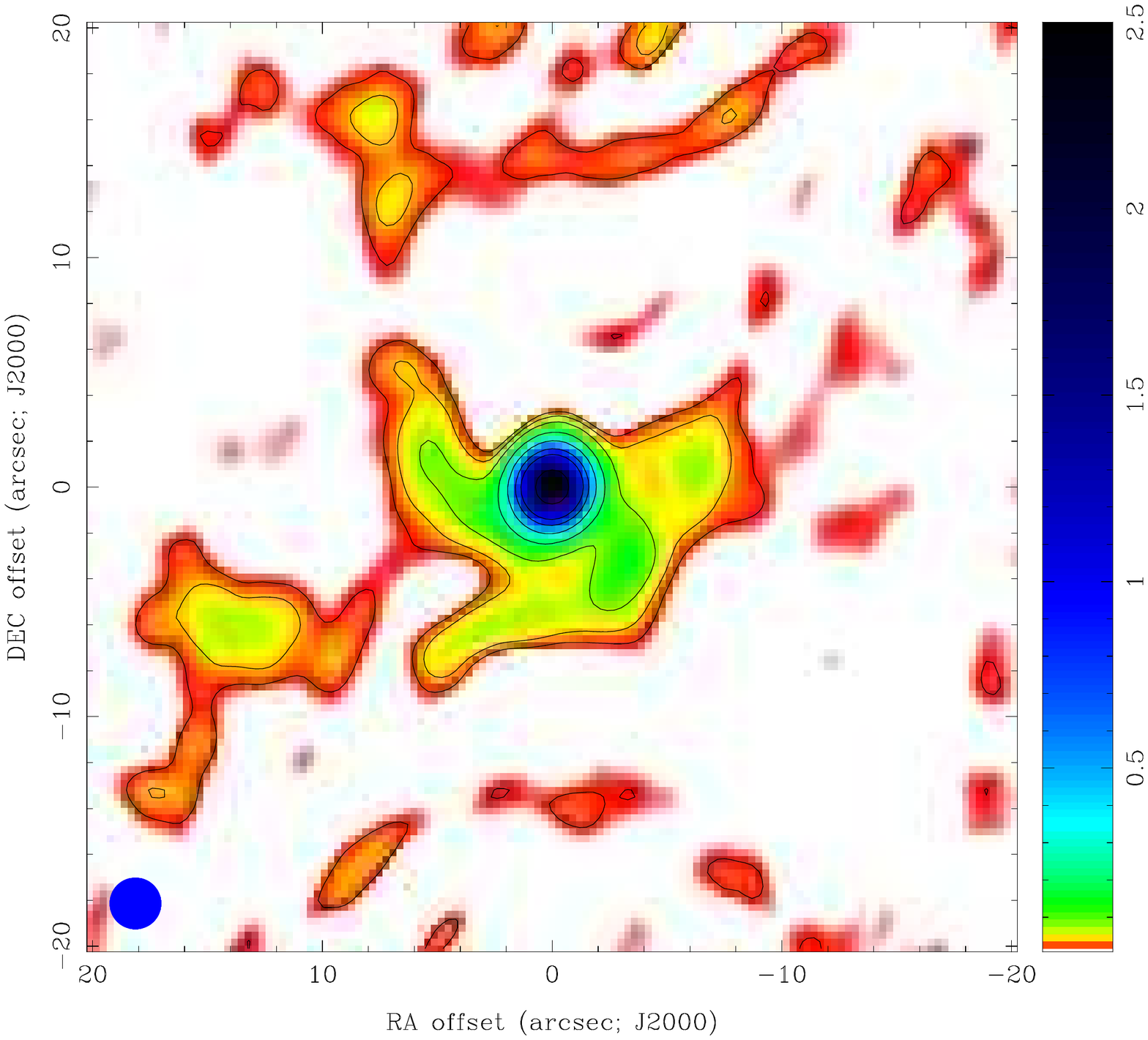}
\caption{\small The same as in Fig. \ref{fig:highres} but with a synthesised circular beam size of 2.2$''$. The different 
resolutions are to highlight emission from the compact and extended features in the region, as explained in Sect. \ref{radiodiscussion}.}
\label{fig:lowres}  
\end{figure*}

\subsection{MIR observations}
The observations at 8.6\,$\mu$m (PAH 1 filter with central wavelength 8.59\,$\mu$m, $\Delta$$\lambda$=0.42\,$\mu$m) 
were carried out in June 2006 with the VLT Imager and Spectrometer for mid-Infrared (VISIR) at the European 
Southern Observatory (ESO) Very Large Telescope (VLT) Unit 3. The field of view of the combined mosaic image 
obtained by shifting and adding background subtracted images obtained by dithering is 
30$''$$\times$30$''$ at a pixel scale of 0.075$''$ \citep{lagage2003}. Further details of the observations, data reduction, 
and flux calibration are given in \cite{schoedel2007}.
% The L- and M-band observations were made in with the ISAAC (Infrared Spectrometer and Array Camera) imager and spectrograph at
% ESO VLT unit telescope UT1 in Chile. The long-wavelength low resolution (LWS3-LR) spectroscopic mode was used with a wavelength 
%range of 2.7\,$\mu$m-4.2\,$\mu$m for L-band observations with the SL filter, and a wavelength range of 4.4\,$\mu$m-5.1\,$\mu$m in 
%the M-band spectral filter. For a 0.6$''$ slit width, this implies a spectral resolution $R=\lambda/\delta\lambda$ of 600 in the L-band 
%and 800 in the M-band. The resulting images are divided by flat fields and corrected for cosmic rays, sky lines and distortion 
%caused by dispersion. A Xenon-Argon lamp is used to calibrate wavelengths.  
\subsection{NIR Brackett$\gamma$}
\label{brgamma}

The Brackett$\gamma$ images were obtained using the NB$\,$2.166 filter of the NAOS/CONICA adaptive optics system at the 
ESO/VLT \citep{lenzen2003,rousset2003}, on August 5th, 2009\footnote{Based on observations made by ESO telescopes at Paranal 
Observatory, under the programme 083.B-0390(A).}.
Standard data reduction, including bad pixel correction, sky subtraction, flat field correction, and the detector-row cross-talk 
correction, was performed using IDL and DPUSER\footnote{developed by T. Ott; \url{http://www.mpe.mpg.de/~ott/dpuser}.}. 

The individual images were combined in a mosaic using a simple shift-and-add algorithm. Shifts between the individual exposures 
were determined by cross-correlation, performed by the $jitter$ routine (part of the ESO $eclipse$ package; \citealt{devillard1997}). 
The images were then shifted and median averaged using DPUSER.
Point source subtraction was performed using $StarFinder$ \citep{diolaiti2000}. Since the FOV of the Br$\gamma$ mosaic is 
significantly larger than the isoplanatic angle at this wavelength ($\sim$$\,$6$''$), we  divided the mosaic into overlapping sub-images. A local PSF extracted from each of the sub-images was used for PSF fitting and subtraction of stars \citep[for a more detailed description of the procedure see][]{schoedel2010,schoedel2010extinct}. 

Residuals at the positions of bright stars are mainly due to imperfections in PSF extraction, and extreme stellar crowding in the 
central parsec of the GC (Fig. \ref{fig:brgamma}). 
The photometric zero-point was calculated using the precise K$_S$-band photometry from \cite{schoedel2010extinct} of 20 stars randomly 
distributed in the field. The K-band magnitudes are then converted to Janskys using the flux density of Vega in the K$_S$-band 
\citep{allen1976}. An offset of $−$0.13 mag was applied to the K$_S$-magnitudes to convert them to Br$\gamma$- magnitudes. This offset 
was calculated using the transmission curves of both the K$_S$ and NB filters and assuming a given extinction law, total extinction, 
and blackbody temperature. The calculated magnitude offset was insensitive to the assumed extinction (which varied between A$_{K_{S}}$ = 2.0--3.5),
 blackbody temperature ($T = 2\times10^3-3\times10^4$\,K), and extinction law (exponent $\alpha$ = 2.0--2.2), and varied 
by $<$0.015 as these parameters were varied. At $\lambda$$\sim$2.30 $\mu$m, the SED of late-type stars deviates strongly from a 
blackbody because of the CO-bandhead absorption dip. However, this effect is limited to a small part of the K$_S$ window and is not 
expected to be stronger than the one causing the blackbody temperature to vary by one order of magnitude. The measured Br$\gamma$-band 
magnitudes were converted into Janskys, assuming 667 Jy for a source of magnitude zero (value for K$_S$ filter from Cox 2000); we 
estimate that the uncertainty in this conversion factor is $<$10\%.  
% \begin{table}[!Ht]
% {\begin{small}
% \begin{tabular}{ccccc}
% \hline
%  $\lambda$ & $B_{major}$ & $B_{minor}$ &PA(degrees)&Flux (Jy)\\
% \hline
%  3.0~mm	 & 2.68 & 1.71	&-20.6&	2.463\\
%  1.3~mm	 & 2.01 & 1.27	&-5.2	&2.491\\
% \hline
% \end{tabular}
% \end{small}}
% \caption{Fluxes of Sgr~A*}
% \label{tablesgr}
% \end{table}

\section{Results}
% \begin{figure*}[!Ht]
% %\begin{minipage}[t]{.4\textwidth}
% \begin{center}
% \includegraphics[scale=0.5,angle=-90]{highres1mm}
% \caption{\small 1mm CD array configuration map with a synthesised circular beam size of 0.9$''$. Contour levels are 0.0065, 
%0.01, 0.02, 0.03, 0.04, 0.05, 0.1, 0.6, 0.9, 1.2, 1.5, 2.5 Jy/beam.}
% \label{fig:highres}  
% \end{center}
% %\end{minipage}  
% \end{figure*}
% \begin{figure*}[!Hb]
% %\begin{minipage}[t]{.4\textwidth}
% \begin{center}
% \includegraphics[scale=0.5,angle=-90]{res19191mm}
% \caption{\small 1mm CD array configuration map with a synthesised circular beam size of 0.9$''$. Contour levels are 0.0065, 0.01, 
%0.02, 0.03, 0.04, 0.05, 0.1, 0.6, 0.9, 1.2, 1.5, 2.5 Jy/beam.}
% \label{fig:lowres}  
% \end{center}
% %\end{minipage}  
% \end{figure*}
\subsection{Radio 3\,mm and 1.3\,mm maps}
\label{radiodiscussion}
We produced a 3\,mm map of resolution 2.68$''$$\times$1.71$''$(P.A.=-20.7$^\circ$) from the C array data of May 2009 
(refer to Fig. \ref{regular3mm}). The primary beam (HPBW) contains a total integrated flux of 13\,Jy. \cite{mezger1989} reported a total 
flux density of 22\,Jy at 106\,GHz with the single dish IRAM 30\,m telescope on Pico Veleta, Spain. The missing flux of $\sim$9\,Jy 
is due to the lower sensitivity of interferometric observations to the underlying extended thermal emission than for the brighter small-scale features of the 
mini-spiral. This is also seen in other interferometric observations, 
such as \cite{shukla2004}, who reported a total flux of 12\,Jy from the central 85$''$ region using the OVRO array at 92\,GHz from a 
6$''$.95$\times$3$''$.47 (P.A.=-5$^\circ$) map, and \cite{wright1987} who reported a total flux of 12.6\,Jy at 86\,GHZ with the Hat Creek 
Interferometer. These agree with our flux measurements. We note that the missing flux does not pose a significant problem for our 
spectral index measurements, as explained in Sect. 3.3.

The uncertainty in the flux density measurements is given by $\sqrt{(\sigma_{rms}^{2})+(\sigma_{cal}F_{\nu})^{2})}$, where 
$\sigma_{rms}$ gives the rms noise of the map, $\sigma_{cal}$ gives the relative error in flux calibration, which we estimate to be 
15\% in our case, and $F_{\nu}$ is the flux density at frequency $\nu$.

The original 1.3\,mm D array map of resolution 4.42$''$$\times$1.23$''$ (P.A.=1.0$^\circ$) is shown in Fig. \ref{regular1mm}. The 
mini-spiral is clearly a complex region with several features of interest. To highlight these features, we present two maps in this 
paper. Fig. \ref{fig:highres} shows the 1.3\,mm CD array configuration map with a synthesized beam size corresponding to  1.2$''$ 
emphasizing the compact structures in the region, which can be clearly distinguished from each other and from the extended emission of the 
mini-spiral. The other map (shown in Fig. \ref{fig:lowres}), with a synthesized beam size corresponding to the lower angular resolution 
of 2.2$''$ allows us to look at the more extended emission. The main features that we can see in the two maps are: the central non-thermal source
 Sgr~A*, and most of the thermal mini-spiral, including the bar, eastern arm and parts of the northern arm. The western arc is resolved 
out in our maps. 
% The fluxes of Sgr~A* at the two different wavelengths obtained from our maps are tabulated in Table \ref{tablesgr}. From the combined 
%C and D array data at 1.3\,mm, we produced a map of resolution 2.01$''$$\times$1.27$''$ (P.A.=-5.2$^\circ$) which is shown in 
%Fig. \ref{regular1mm} in the Appendix.

\begin{figure}[!Htp]
%\begin{minipage}[t]{.4\textwidth}
 \begin{center}
\includegraphics[scale=0.5,angle=0]{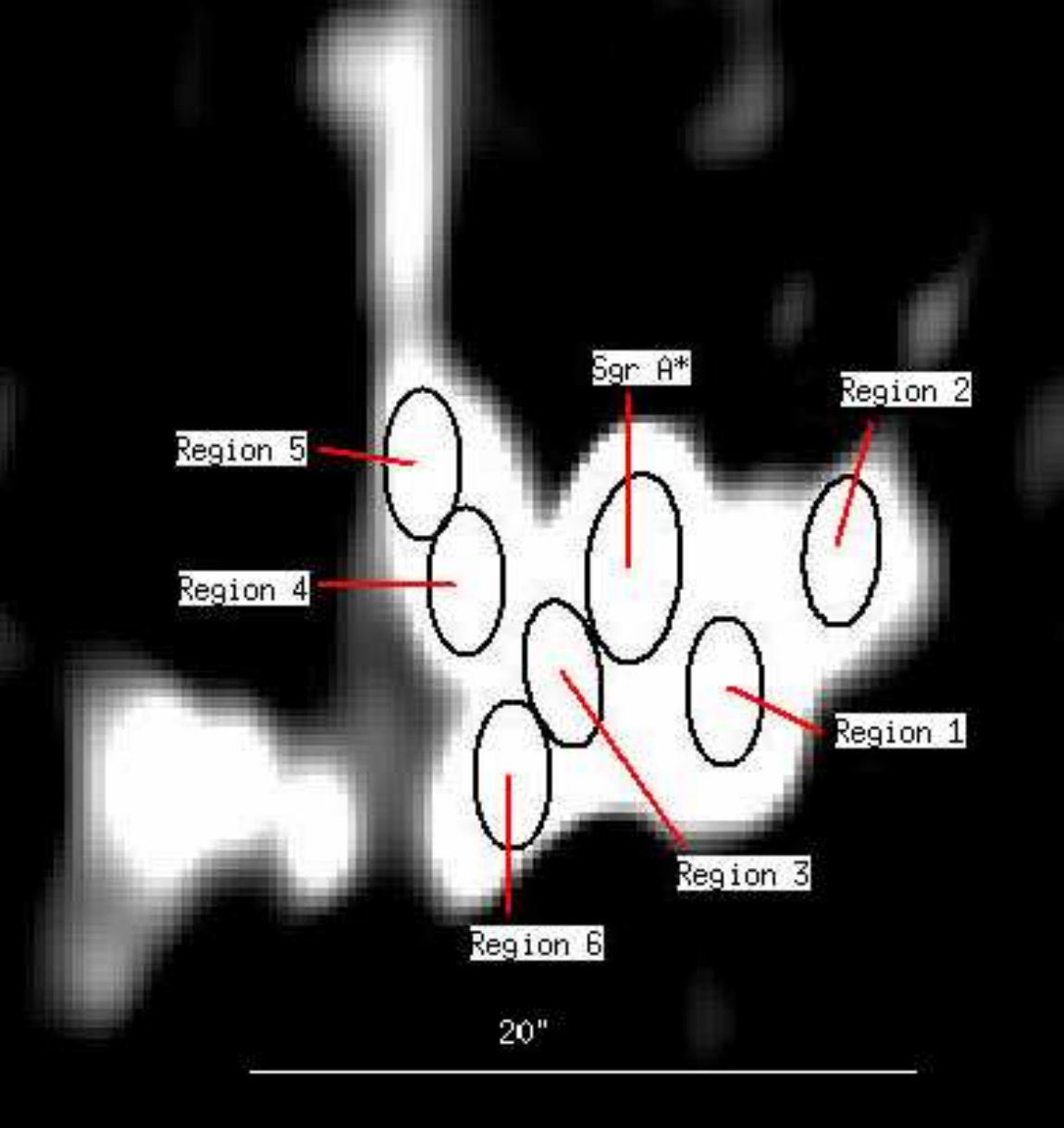}
\caption{\small \small The elliptical regions used to extract fluxes from the central 20$''$ region of the GC from the multi-wavelength 
datasets, marked on the 1.3\,mm radio continuum map of resolution 4$''$$\times$$2''$ (P.A=0$^\circ$).}
\label{fig:regions}  
 \end{center}
%\end{minipage}  
\end{figure}

To compare fluxes in different maps at different wavelengths, we chose six regions in the point-source-subtracted 
version of the extended emission map. Since the beam shape is elliptical, we chose elliptical regions (Fig. \ref{fig:regions}). 
The elliptical shape also allows us to fit the elongated structures in the mini-spiral region more closely, hence maximising the aperture area 
on source and avoiding any loss of flux from the source. The fluxes extracted from these sources in the 1.3\,mm and 3\,mm maps are tabulated in 
Table \ref{fluxtable}. Using the kinetic temperatures estimated by \cite{zhao2010} in the mini-spiral arms from radio recombination 
lines (H92$\alpha$ and H30$\alpha$), we calculated the properties of the selected regions including the emission measure and electron 
density from the free-free radio continuum emission flux densities at 3\,mm \citep{panagia1978}. We measured electron densities in the range 0.8-1.5$\times$10$^4$\,cm$^{-3}$ in these regions, with the highest electron densities and emission occurring in the IRS 13 region. 
\cite{zhao2010} obtained slightly higher values of number densities of 3--21$\times$10$^4$\,cm$^{-3}$ for the same regions. 
However, our values are comparable to values of number densities of the order of 2$\times$10$^4$\,cm$^{-3}$ obtained from radio 
continuum measurements in \cite{brown1981} and \cite{lo1983} and from Pa$\alpha$ measurements in \cite{scoville2003}. We can conclude that the 
derived values are consistent with previously published results within the errors.

% At 3\,mm, the mini-spiral region is believed to consist of mainly free-free emission. Using this assumption, we scale our 3\,mm 
%map to 1\,mm using the spectral index notation (S$\sim$$\nu^{-0.1}$), and subtract this scaled map from our 1\,mm map. The residual 
%map indicates positive flux in certain regions of the mini-spiral, mainly in the Eastern Arm, which we believe to be the dust emission 
%contribution beginning to show up. 
\subsection{Spectral index}
\label{spindex}
Our spectral index map \citep[to be published in the upcoming conf. proc.][]{kunneriath2012} covers the inner 40$''$$\times$40$''$ of the GC region, which includes the 
northern arm, eastern arm, the bar and the central bright radio source Sgr A* and was produced from 
the 3 and 1.3\,mm images.

We used the  notation S$_{\nu}$ $\sim$ $\nu$$^{\alpha}$, 
where $\alpha$ is the spectral index and S$_\nu$ is the 
flux density at frequency $\nu$. We convolved our 3\,mm and 1.3\,mm maps to an angular resolution of 4$''$$\times$2$''$, after which 
a primary beam correction was applied. We applied a 
flux density cutoff of 0.01\,Jy to both the images to exclude regions of low flux density to create masked 
images at both frequencies. By multiplying these masked images, we created a final mask, which was then used 
to obtain the spectral index map of the region. The masking process ensures that the spectral index is well-defined at all points in 
the masked region by avoiding any division by low flux values. With the calculated flux density calibration accuracy $\sim$15\%, we 
measured the uncertainty in the spectral index values to be $\delta$$\alpha$$\sim$0.25 from 
$\delta$$\alpha$=1/ln($\nu$1/$\nu$2)$\times$$\sqrt{({\delta}S1/S1)^{2}+({\delta}S2/S2)^{2}}$, derived from the 
spectral index equation.\\
% \begin{figure}[!Ht]
%     \begin{center}  
%       \includegraphics[scale=0.4,angle=-90]{spinnew}
%       \caption{Spectral index map produced from the 3 and 1.3\,mm maps at a resolution of 4$''$$\times$2$''$ (P.A.=0$^\circ$). Contours 
% indicate spectral indices of -0.1 (dashed), 1.0, and 0.4. Sgr~A* has an inverted synchrotron spectral index of 0.5. The ellipse indicates the beam size.}
%       \label{fig-2}
%     \end{center}
% \end{figure}
% \noindent
% \begin{figure}[!Ht]
% %\begin{minipage}[t]{.4\textwidth}
% \includegraphics[scale=0.45,angle=-0]{onemmexcesscolorbarwithbar}
% \caption{\small 1.3\,mm excess emission (compared to Bremsstrahlung) at a resolution of 4$''$$\times$2$''$ (P.A.=0$^\circ$). See 
%text for details of subtraction of free-free emission. The contour levels represent -0.01,0.01 and 0.05 Jy/beam.}
% \label{fig:3}  
% %\end{minipage}  
% \end{figure}
% \begin{figure*}[!Hb]
% \begin{minipage}{0.5\linewidth}
% \centering
% \includegraphics[scale=0.4,angle=-90]{mmmiroverlay1}
% \caption{\small MIR 8.6\,$\mu$m map overlaid with a 3\,mm map of synthesised circular beam size of 1$''$. Contour levels are at 
%0.0090, 0.015, 0.02, 0.035, 0.3, 0.4, 0.45 Jy/beam.}
% \label{fig:mirmm}
% \end{minipage}%
% \begin{minipage}{0.5\linewidth}
% \centering
% \includegraphics[scale=0.4,angle=-90]{ratiodust}
% \caption{\small Dust to gas ratio map with a synthesised circular beam size of 1$''$ - High at IRS 10W, and IRS 1W (Northern Arm), 
%low at IRS 13 (Bar)}
% \label{fig:dustgas}
% \end{minipage}
% \end{figure*}

We obtained a spectral index of $\sim$0.5$\pm$0.25 for Sgr A*, indicating an inverted synchrotron spectrum, 
which agrees with the results of \cite{falcke1998} and \cite{zadeh2006b}. 
The mini-spiral region around Sgr A* shows a mixture of positive and negative spectral indices.
Free-free thermal bremsstrahlung emission is indicated by the $\sim$$-$0.1 spectral index value, around the bar and parts of the 
northern arm. The steeper spectral indices of $\sim$$-$0.5 could be a result of unresolved flux in the 1.3\,mm map, 
in particular in the northern arm of the mini-spiral. The positive indices in the spectral index map in the mini-spiral 
region, around the edges of the bar and most of the eastern arm, on the other hand, could be indicators of dust emission that 
starts to become significant at wavelengths $\le$1.3\,mm. 
% Since the positive spectral index indicates excess emission in the 1.3\,mm, and 
% assuming the 3\,mm emission largely arises from thermal bremsstrahlung, we scale the 3\,mm map to 1.3\,mm by a factor 
% 0.92 (obtained from the power law flux relation S$\sim$$\nu^{\alpha}$), and subtract this from the 1.3\,mm map. As the bright Sgr\,A* 
%dominates the spectral index in the central region, we mask the central region to exclude the emission from Sgr A*. The resultant 
%residual map (Fig. \ref{fig:3}) shows evidence 
% of excess emission (compared to Bremsstrahlung) in the regions corresponding to positive spectral index regions in the spectral index 
%map, the Eastern Arm, parts of the Bar, and a portion of the Northern Arm. 
% We attribute this excess to the dust emission, which starts to become important at wavelengths$\leq$1\,mm, possibly from large dust 
%grains. Although the overall dust temperature in the mini-spiral region is $\sim$300\,K, lower temperatures and larger grain sizes 
%would be required to explain the excess emission at mm-wavelengths \citep{hildebrand1983,fich1991}. 
The regions of excess emission approximately coincide with locations of marginally cooler (at least by 10\,K) dust, as seen in the 
colour temperature map derived from 12.5\,$\mu$m and 20.3\,$\mu$m maps by \cite{cotera1999}, who found that the overall temperature in the 
mini-spiral is of the order of $\sim$200\,K. This indicates that we may have an unspecified amount of dust (of unknown mass and 
filling factor) at temperatures between 30\,K and 200\,K, with larger-sized dust grains. \cite{moultaka2009}, who found evidence of 
intrinsic CO absorption at 4.6\,$\mu$m in the mini-spiral region, suggest that the presence of lower temperature material in the 
region is not inconsistent with the higher overall temperature, since compact dusty structures with high optical depths can have 
travel times of the order of 10$^3$ years, which is comparable to photo-evaporation timescales of 10$^3$-10$^5$ years, allowing the 
cooler material to travel through the central parsec. 

\subsection{Uncertainty in spectral index}
Spectral index maps are very sensitive to zero-spacing flux density offsets, in particular for radio interferometric observations, 
and if the observed wavelengths are as close as in our case. However, by choosing the D configuration at 1.3\,mm we cover the range 
in angular resolution obtained in the C configuration at 3\,mm very well. In the C configuration at 100\,GHz, the angular resolution 
ranges from 2.16$''$ for the shortest baseline of 30\,m to 25.16$''$ for the longest baseline of 350\,m, while in the D configuration 
at 230\,GHz, the angular resolution ranges from 2.18$''$ for the shortest baseline of 11\,m to 29.74$''$ for the longest baseline of 
150\,m. Thus the effects of missing flux at short spacings can be assumed to be similar at both frequencies. The uv-coverages at both 
frequencies, given in the Appendix, demonstrate this similarity.

The spectral index map derived from this data contains valuable information about the flux density differences between both observing 
frequencies. This can be discussed in terms of emission mechanisms and resolution effects.  The similarities between the maps indicate that 
we detect mainly emission from the compact components from within the mini-spiral, which has been successfully 
detected by all mm-interferometers (see references in Sect. \ref{radiodiscussion}). We found that the median spectral index we obtained 
for these compact features is $-$0.1, in full agreement with free-free emission. If the sources suffered an increasing amount 
of resolution on the CARMA baselines we would expect much steeper spectral index values. This is not observed, with the exception of 
the northern arm (see Sect. \ref{spindex}).

\subsection{MIR map}
\begin{figure}[!Hb]
\centering
\includegraphics[scale=0.37,angle=0]{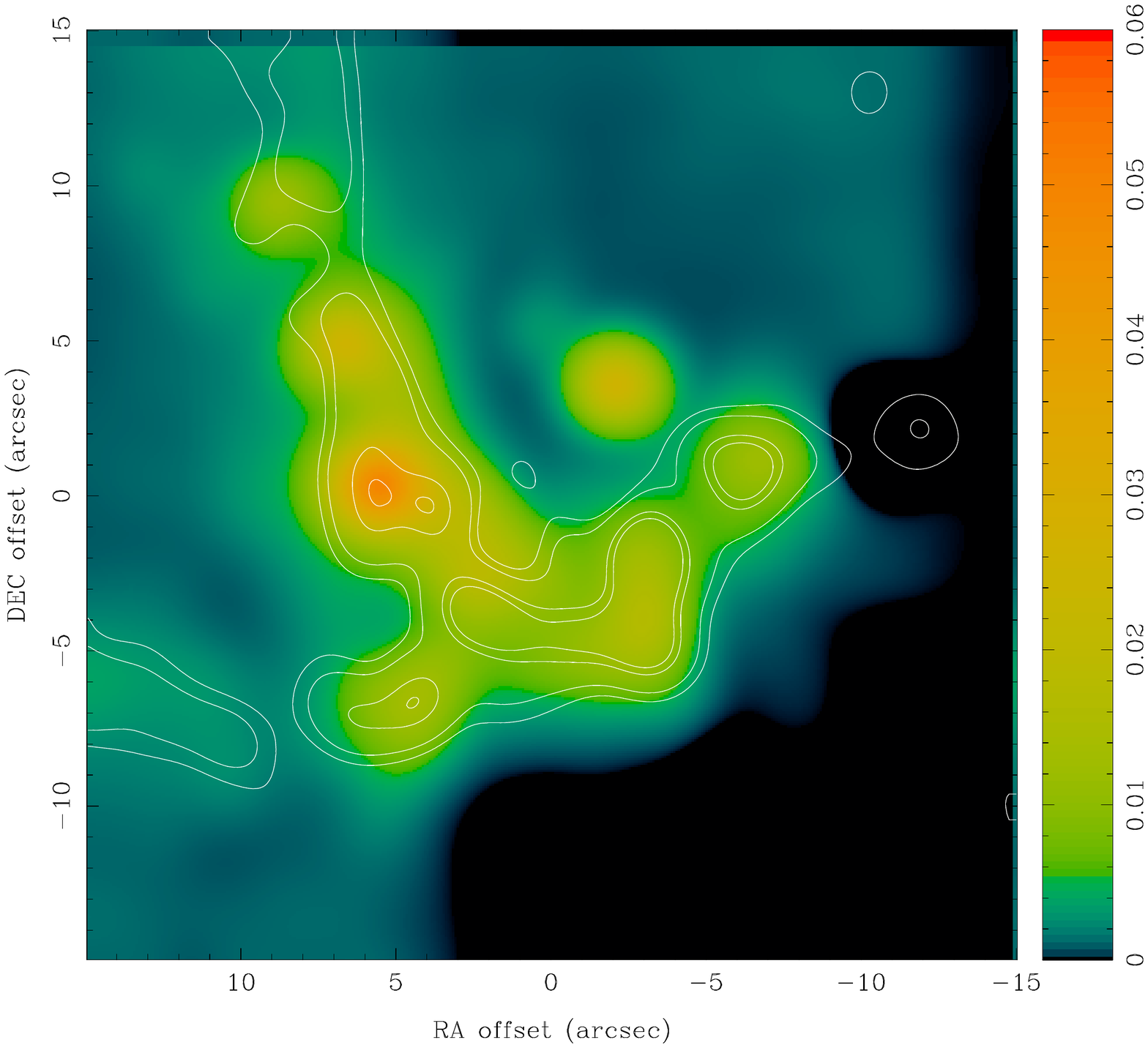}
\caption{\small MIR 8.6\,$\mu$m map overlaid with a 3\,mm point source subtracted map of synthesised circular beam size of 2$''$. 
Contour levels are at 0.02,0.035,0.08,0.1, and 0.2 Jy/beam.}
\label{fig:mirmm}
\end{figure}

The MIR map at 8.6\,$\mu$m reveals extended dust emission in the central parsec region. A strong correlation was observed between 
the warm dust emission at MIR (12.4\,$\mu$m) wavelengths and the ionized gas emission at radio wavelengths (VLA 2\,cm) by 
\cite{gezari1991}, indicating that the dust and gas are mixed in the region, with the point sources IRS 1, 2, and 9 coincident in 
both maps and significant displacements in the position of IRS 13 and 21 between the two which could be due to dust displacement 
produced by stellar winds. An overlay of our high resolution C array 3\,mm map over the 8.6\,$\mu$m map (at 2$''$ resolution) is shown 
in Fig. \ref{fig:mirmm}, which confirms this correlation, with the radio contours tracing out the dust in the MIR quite well.

The uncertainties in flux density measurements are obtained from the rms error and flux calibration error (estimated to be $<$10\%). 
Fluxes from the same regions selected in the radio continuum maps are extracted. The dust masses for the regions are calculated using 
\begin{equation}
M_d=\frac{F(\nu)D^2}{B(\nu,T_d)}\frac{4a}{3Q(\nu)}\rho,
\end{equation}
where $F(\nu$) is the flux density, $D$ is the distance to the region, $a$, $Q(\nu)$, and $\rho$ are the dust parameters grain size, 
emissivity, and grain density, respectively, and $B(\nu,T)$ the Planck function at dust temperature $T_d$, which we assume to be 
200\,K \citep{cotera1999}. Dust grain parameters were obtained from \cite{rieke1978}. The masses are tabulated in Table \ref{fluxtable}. 
  
\subsubsection{Dust-to-gas ratio}
\begin{figure}
\centering
\includegraphics[scale=0.37,angle=0]{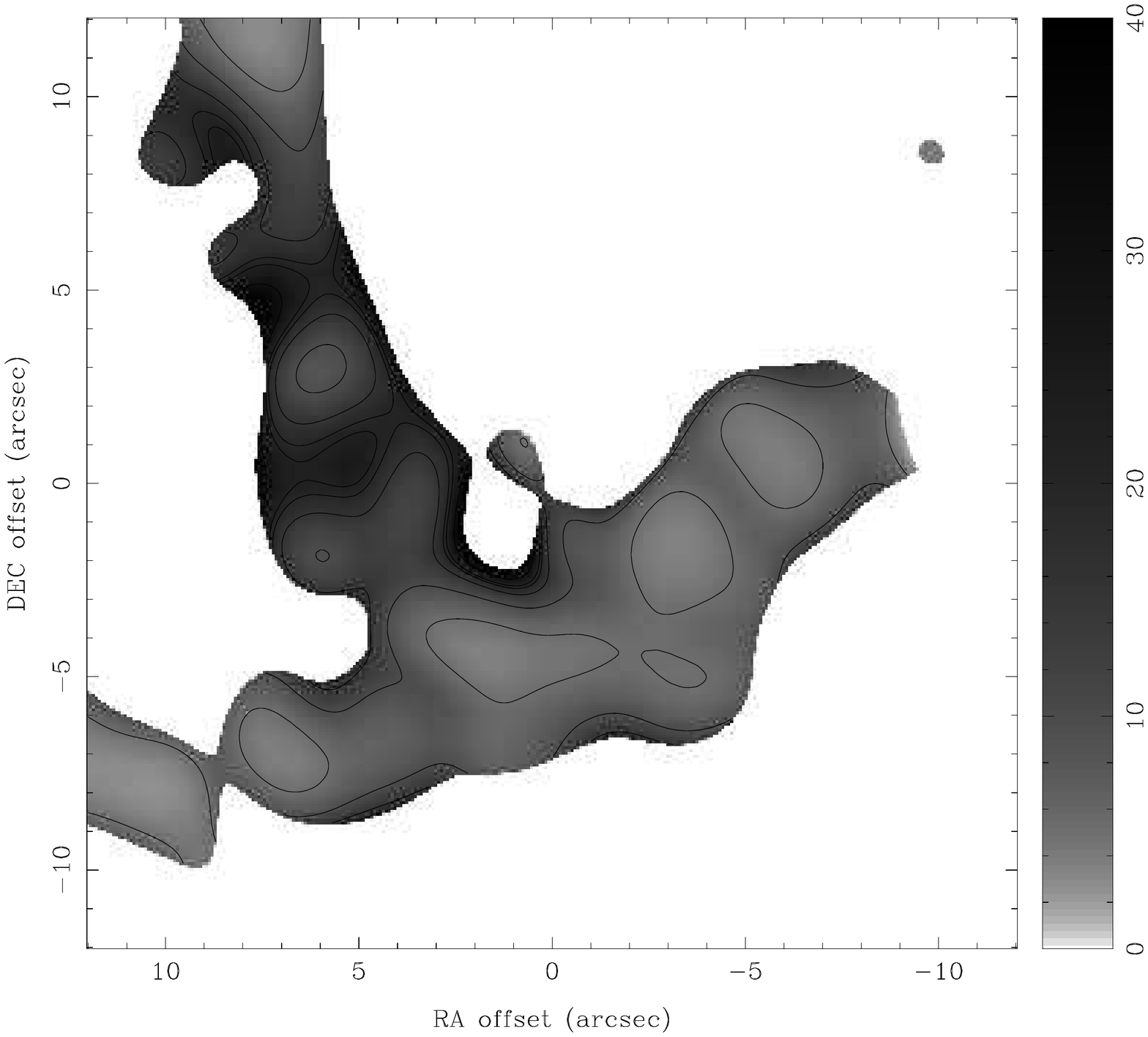}
\caption{\small Dust to gas ratio map with a synthesised circular beam size of 2$''$ - High at IRS 10W, and IRS 1W (Northern Arm), 
low at IRS 13 (Bar).}
\label{fig:dustgas}
\end{figure}

The 3\,mm extended emission map and the 8.6\,$\mu$m map were converted to the same pixel scale and shifted to match each other 
\citep[using the position of Sgr~A* in the MIR from][]{schoedel2007}, to within an accuracy of 0.075$''$ and used to obtain the 
dust-to-gas brightness ratio map shown in Fig. \ref{fig:dustgas}. The dark regions in the map indicate higher dust-to-gas ratios, 
while the lighter regions corresponds to less dust, with the values ranging from $\sim$10 in the IRS 13 region to $\sim$30 in the 
IRS 1W region. The higher ratios in the northern arm than in the bar are consistent with the explanation that the ionized bar 
of the mini-spiral region is dust depleted by an outflow/wind from the central ionizing source Sgr~A* \citep{gezari1991}.

% \begin{figure}
% %\centering
% \includegraphics[scale=0.5]{radiomir}
% \caption{\small Overlay map of radio and MIR}
% \label{fig:4}    
% \end{figure}
% \begin{figure}[Htbp]
% %\centering
% \includegraphics[scale=0.4,angle=-90]{testdust}
% \caption{\small Dust to gas ratio map - High at IRS 10W, and IRS 1W (Northern Arm), low at IRS 13 (Bar)}
% \label{fig:4}    
% \end{figure}
\subsection{Brackett$\gamma$ map}
The Br$\gamma$ hydrogen recombination line at 2.16\,$\mu$m is a useful indicator of the ionized gas in the mini-spiral region, which traces 
out the same ionized emission seen in the radio continuum maps at 3\,mm. \cite{neugebauer1978} report that most of the background flux 
at 2.2\,$\mu$m is from the stars in the region and estimate the free-free emission contribution to the continuum radiation at 
2.2\,$\mu$m to be negligible. Therefore, after removing the strong stellar sources using a point-source subtraction method described 
in Section \ref{brgamma}, we derived a reliable estimate of the emission line fluxes in the region. 
\begin{table*}[!Ht]
\centering
{\begin{small}
\begin{tabular}{c|c|c|c|c|c|c|c|c|c}
\hline
Region &  $S_{\mathrm{3{\,}mm}}$ & $S_{\mathrm{1.3{\,}mm}}$ & $S_{\mathrm{8.6{\,}{\mu}m}}$ & $S_{\mathrm{Br\gamma}}$ & $EM$	& $n_{e}$	& $N_{\mathrm{Lyc}}$	& $M_{\mathrm{dust}}$ & $\alpha$	\\
 & Jy & Jy & Jy & Jy &  $\times$$10^6$$\mathrm{cm}^{-6}$$\mathrm{pc}$ & $\times$$10^4$$\mathrm{cm}^{-3}$	&  $\times$$10^{50}$$\mathrm{s}^{-1}$ & \,\solm&3 to 1.3\,mm\\
\hline
& & & & & & & & &\\
&$\pm$0.03 & $\pm$0.03 & $\pm$5 & $\pm$0.015 & $\pm$0.8 & $\pm$0.3 & $\pm$0.6 & $\pm$0.006&\\
\hline
& & & & & & & & &\\
1&	0.24    &0.22	&	20.9	&0.096	&9.71	&1.45	&3.43	&0.016&$0.02_{-0.03}^{+0.08}$\\
& & & & & & & & &\\
2&	0.12    &0.10	&	16.2	&0.042	&2.02	&0.95	&2.08	&0.013&$-0.09_{-0.05}^{+0.04}$\\
& & & & & & & & &\\
3&	0.15	&0.15	&	23.6	&0.080	&6.07	&1.15	&2.57	&0.018&$-0.16_{-0.08}^{+0.03}$\\
& & & & & & & & &\\
4&	0.14	&0.13	&	46.6	&0.125	&3.43	&1.05	&2.48	&0.036&$-0.07_{-0.09}^{+0.14}$\\
& & & & & & & & &\\
5 &	0.08	&0.07	&	21.6	&0.062	&1.99	&0.79	&1.48	&0.017&$-0.12_{-0.02}^{+0.05}$\\
& & & & & & & & &\\
6&	0.09	&0.07	&	16.6	&0.049	&4.09	&0.89	&1.79	&0.013&$-0.10_{-0.06}^{+0.06}$\\
& & & & & & & & &\\
%7&	0.065$\pm$0.0018	&0.088$\pm$0.0019	&	9.760	&0.056	&1.75	&0.82	&1.52	&0.0005\\
%& & & & & & & &\\
\hline
\end{tabular}
\tablefoot{Fluxes from different wavelengths, calculated emission 
measure, electron density, number of ionizing photons, dust mass, and spectral index (with its variation across the region) for selected 
regions from the mini-spiral. Positions of the 
selected regions are shown in Fig. \ref{fig:regions}. The second row gives the uncertainties in values.}
\end{small}}
\caption{Fluxes and calculated properties of selected mini-spiral regions}
\label{fluxtable}
\end{table*}
Similarities between the observed structures of the emission at Br$\gamma$ and 3\,mm wavelengths suggest that they largely trace emission 
from the same compact sources in the region. This allows us to use the CARMA 3\,mm map to derive 5\,GHz continuum and Br$\gamma$ 
line flux maps, respectively, as described below.

\begin{figure}[!Htbp]
%\centering
\includegraphics[scale=0.37,angle=0]{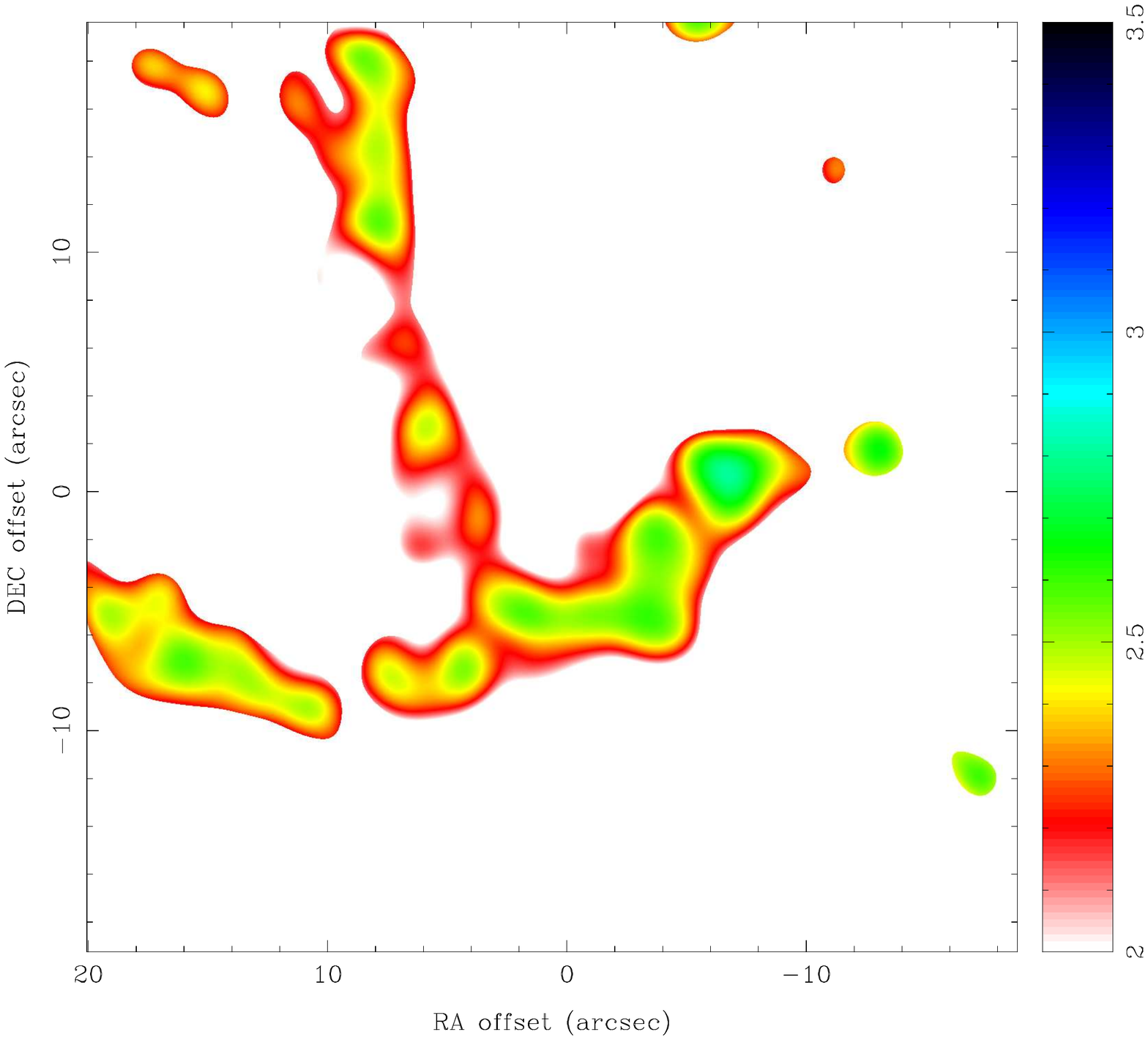}
\caption{\small Extinction map at 2.16\,$\mu$m (synthesized circular beam size of 2$''$).}
\label{extinct}    
\end{figure}

Assuming an optically thin free-free emission spectral index of $-$0.1 between 5\,GHz and 100\,GHz, we scaled our 100\,GHz CARMA 
map to obtain a 5\,GHz flux density map. The expected Br$\gamma$ line intensity can then be calculated from the 5\,GHz radio 
continuum flux density using the formula \citep{glass1999}
\begin{equation}
I(Br\alpha)=2.71\times10^{-14}\left(\frac{T}{10^4\,K}\right)^{-0.85}\left(\frac{\nu}{GHz}\right)^{0.1}F_{\nu}(\rm{Jy})\;\;\;\rm{Wm^{-2}}
\end{equation}
From the ratio of observed to expected flux density, we obtained an estimate of the extinction in the mini-spiral region at 2.16\,$\mu$m. 
The extinction ranges from 1.8--3.0\,mag in the mini-spiral arms, with a mean extinction of 2.30$\pm$0.16\,mag. This is in good 
agreement with \cite{schoedel2010}, who reported a median extinction value of A$_{Ks}$=2.74$\pm$0.30\,mag, using H-K colours. It is also 
in good agreement with \cite{brown1981} and \cite{scoville2003}, who derived extinction values from the radio continuum emission 
and the Br$\gamma$ and Pa$\alpha$ line emission, respectively. Fig. \ref{extinct} gives the extinction map at 2.16\,$\mu$m. 
%This corresponds to an A$_v$ ranging from 25-32$\pm$2.0, using $A_{2.2\mu m}$/$A_v$=0.105 \citep{cardelli1989}. 
Following \cite{ho1990}, the number of ionizing photons was inferred from the expected Br$\gamma$ flux densities, which are tabulated 
in Table \ref{fluxtable}.   
\section{Summary}
We have studied the mini-spiral at the Galactic centre at multiple wavelengths, including radio continuum maps at 
3 and 1.3\,mm, MIR continuum at 8.6\,$\mu$m, and the Br$\gamma$ line emission at 2.16\,$\mu$m.  
We have presented high resolution maps of both this mini-spiral and Sgr~A* at 100\,GHz and 230\,GHz. 
The spectral index map indicates that a mixture of 
various emission mechanisms operate in the central few parsecs of the GC, with an inverted synchrotron spectrum of $\sim$0.5 of Sgr~A*, thermal 
free-free emission spectral indices of $-$0.1, and a possible contribution of dust emission as indicated by the positive thermal 
indices $\sim$1.0. We attributed the positive spectral indices to a possible dust contribution that is evident at wavelengths 
$\le$1.3\,mm, probably from large dust grains \citep{draine1984}.  

For selected regions in the mini-spiral arms, we extracted flux densities from the different wavelengths and inferred the following:
\begin{itemize}
\item The physical properties of the ionized gas such as emission measure are 2--10$\times$$10^6$cm$^{-6}$pc and the electron densities 
are 0.8--1.5$\times$10$^4$\,cm$^{-3}$ according to the 3\,mm radio continuum map. 
\item The dust masses inferred from the 8.6\,$\mu$m MIR map, with a total dust mass contribution of $\sim$0.25$M_{\sun}$ from the mini-spiral 
arms, and a dust-to-gas brightness ratio map indicating dust depletion along the bar, with higher ratios along the northern arm. 
\item We inferred extinctions of 1.8--3.0 at 2.16\,$\mu$m from a comparison of observed Br$\gamma$ flux densities to the expected flux densities 
inferred from the free-free radio emission and the Lyman continuum emission rate derived from the expected Br$\gamma$ flux density.   
\end{itemize}

\section*{Acknowledgements}
D. Kunneriath and M. Valencia-S. are members of the International Max Planck Research School (IMPRS) for 
Astronomy and Astrophysics at the MPIfR and the Universities of 
Bonn and Cologne. RS acknowledges support by the Ram\'on y Cajal
programme by the Ministerio de Ciencia y Innovaci\'on of the
government of Spain. Macarena Garcia-Marin is supported by the German federal department for 
education and research (BMBF) under the project numbers: 50OS0502 \& 
50OS0801. Part of this work was supported by the COST Action MP0905: Black Holes in a violent Universe and 
PECS project No. 98040.
This research has made use of NASA's Astrophysics Data System.

\section*{Appendix}
\begin{figure}[!Htbp]
\centering
\includegraphics[scale=0.7]{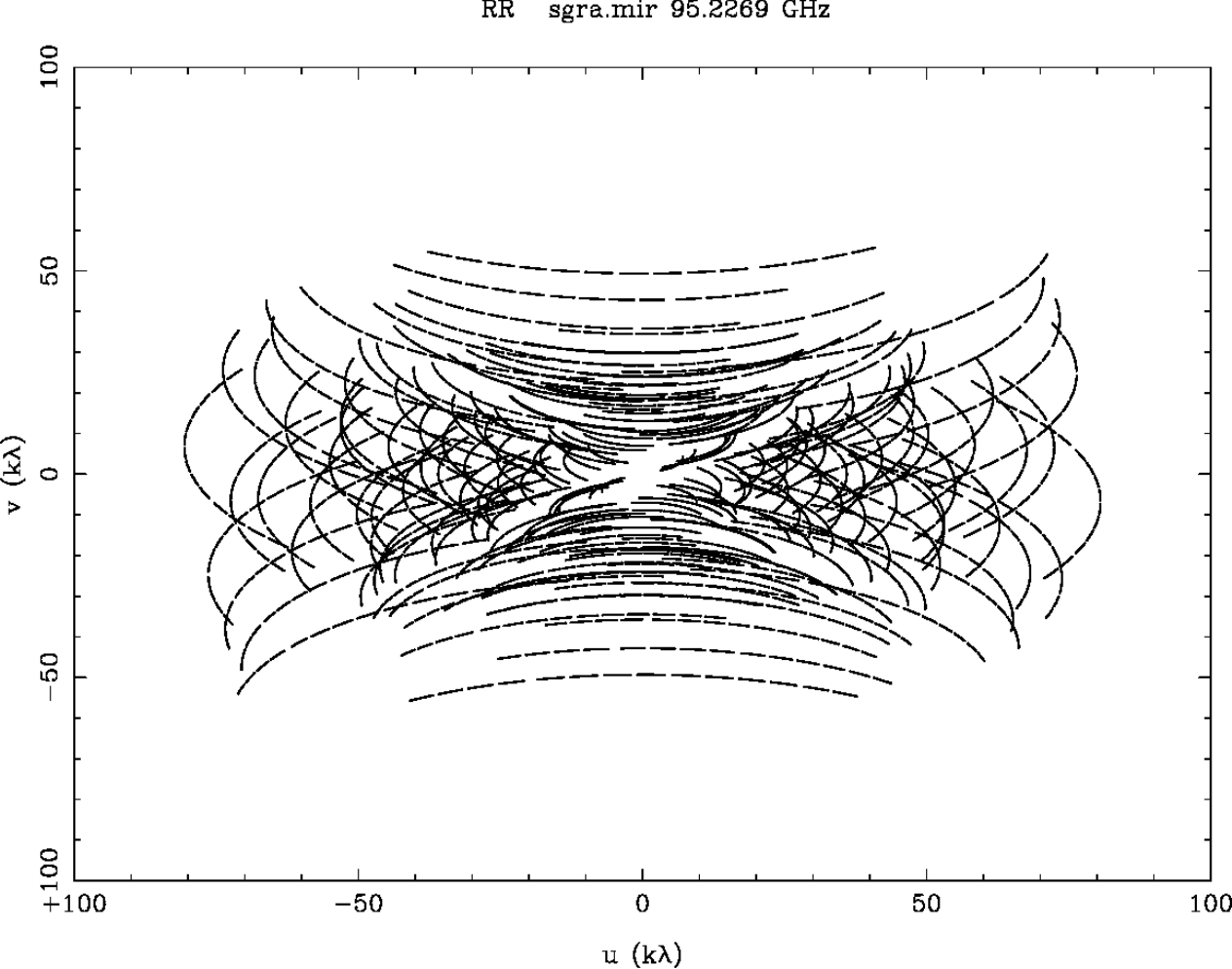}
\caption{\small C array UV coverage at 3\,mm}
\label{uv1mm}    
\centering
\includegraphics[scale=0.7]{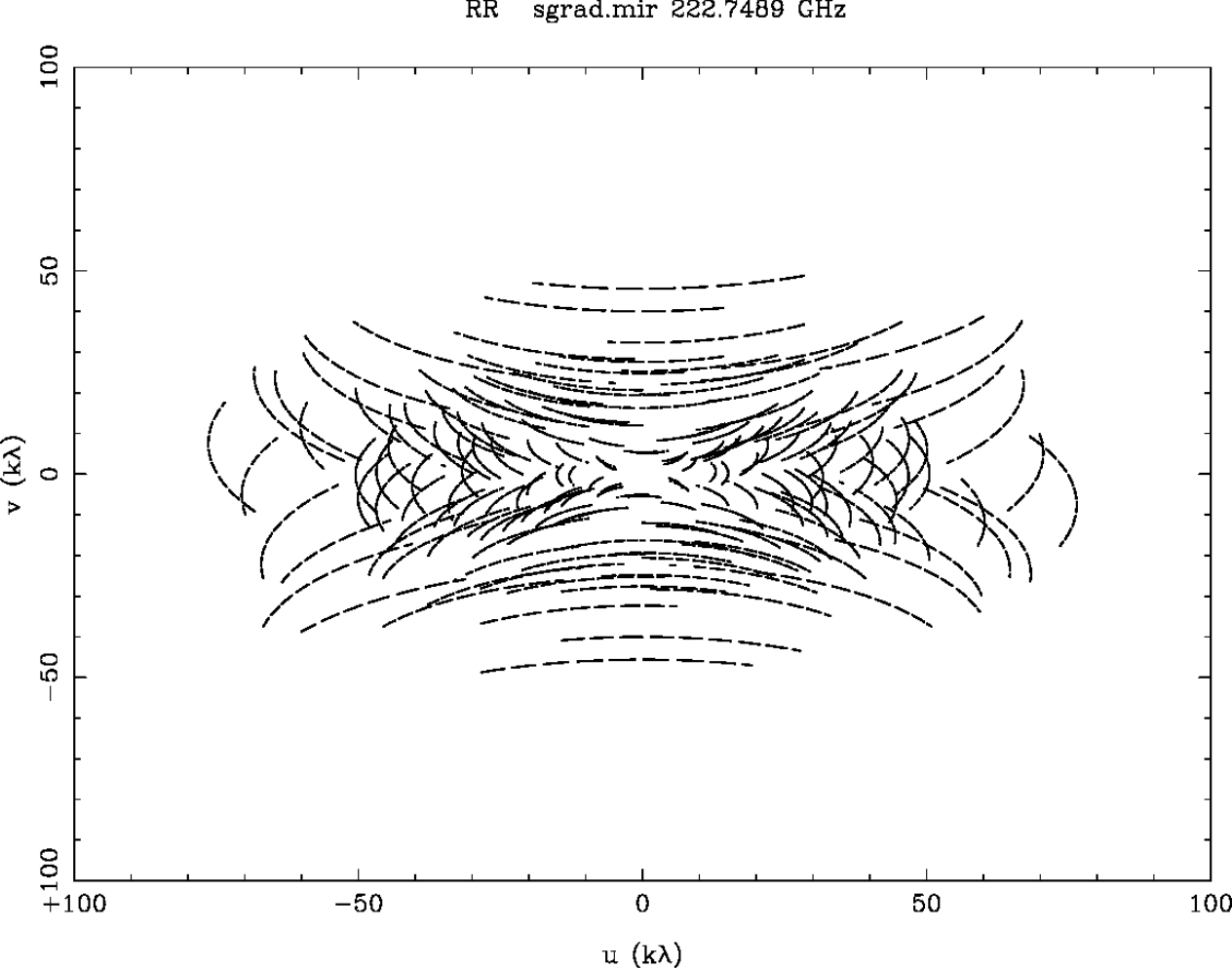}
\caption{\small D array UV coverage at 1.3\,mm}
\label{uv3mm}    
\end{figure}

Figs. \ref{uv1mm} and \ref{uv3mm} show the uv-coverage of the C array at 3\,mm and the 
D array at 1.3\,mm. To derive reliable maps 
and determine the spectral index from the interferometric data, the uv-coverage of the maps 
must match each other very closely at different wavelengths. Since they correspond to different 
configurations of the CARMA array at different wavelengths, they cannot be exactly identical. 
However, they are comparable in uv-distance and appear to be similar enough to produce maps of 
comparable fluxes, using the data 
reduction method described in Section 2.1.

\bibliography{mybibminispiral}{}
\bibliographystyle{aa}

\end{document}